\documentclass[%
 reprint,
superscriptaddress,
nofootinbib,
 amsmath,amssymb,
 aps,
floatfix,
twocolumn,
pre]{revtex4-2}

\usepackage{xcolor}
\usepackage{comment}
\usepackage{graphicx}
\usepackage{dcolumn}
\usepackage{bm}
\usepackage{upgreek}

\begin{document}

\preprint{APS/123-QED}

\title{Training and re-training liquid crystal elastomer metamaterials for pluripotent functionality
}

\author{Savannah D. Gowen}
 \email{savannah.gowen@mat.ethz.ch}
 \affiliation{
 Department of Physics and The James Franck and Enrico Fermi Institutes,
 University of  Chicago, Chicago,  IL  60637,  USA}
\author{Elina Ghimire} 
    \affiliation{Pritzker School of Molecular Engineering, University of  Chicago, Chicago,  IL  60637,  USA}  
\author{Charlie A. Lindberg} 
    \affiliation{Pritzker School of Molecular Engineering, University of  Chicago, Chicago,  IL  60637,  USA}\author{Ingrid S. Appen} 
    \affiliation{Pritzker School of Molecular Engineering, University of  Chicago, Chicago,  IL  60637,  USA}  
    
\author{Stuart J. Rowan} 
     \affiliation{Pritzker School of Molecular Engineering, University of  Chicago, Chicago,  IL  60637,  USA}  
    \affiliation{
 Department of Chemistry, 
 University of  Chicago, Chicago,  IL  60637,  USA}

\author{Sidney R. Nagel} 
    \affiliation{
 Department of Physics and The James Franck and Enrico Fermi Institutes,
 University of  Chicago, Chicago,  IL  60637,  USA}

\date{\today}

\begin{abstract}
Training has emerged as a promising materials design technique in which function can be achieved through repeated physical modification of an existing material rather than by direct chemical functionalization, cutting or reprocessing. This work investigates both the ability to train for function and then to erase that function on-demand in macroscopic metamaterials made from liquid crystal elastomers (LCEs). We first show that the Poisson’s ratio of these disordered arrays can be tuned via directed aging to induce an auxetic response. We then show that the arrays can be reset and re-trained for another local mechanical function, allostery, thus demonstrating pluripotent functionality.  
\end{abstract}

\maketitle


\section{Introduction}
Focused training often leads to stored memories~\cite{learning}. Learning in the brain can occur by repeated exposure to input stimuli; likewise, “muscle memory” is the result of a training regimen that targets specific physical performance~\cite{muscle}. Over time, training can improve the performance of specific mental or physical tasks. However, forgetting previous training is also often advantageous for acquiring new, perhaps incompatible, skills~\cite{forgetting}.

The abilities to remember, forget, and train are not properties exclusively concerned with biological systems~\cite{training,sternlearning,dillavou}; even ordinary physical materials exhibit various forms of memory as a result of training~\cite{keimmemory,nagelmemory}. Training is unlike material design techniques where the microscopic or structural features responsible for function are incorporated at the time of fabrication. Instead, function is achieved through the modification of existing material properties in response to external stimuli; for example, via extended or repetitive exposure to strains involved in a desired mechanical function. However, some materials can be trained more easily than others, and indeed, some materials may be incapable of being trained at all. As such, more work is necessary to determine what materials are trainable and why.

Less consideration has been given to materials that can both train and forget. Admittedly, forgetfulness has often been presented as an obstacle rather than a feature~\cite{keim_forget} . However, the ability to forget introduces the possibility of creating materials that can first be trained for one function and then, when there is a need, retrained for another. In this work, we examine the ability of macroscopic disordered geometric arrays made from liquid crystal elastomers (LCEs) to train in one function and then to forget it on-demand.

Disordered macroscopic arrays, composed of nodes connected by struts, are model materials for many elastic systems~\cite{training}. Simulations and experiments have demonstrated that such structures can be trained to exhibit novel bulk and local properties~\cite{nidhi_aging,hexner_creep,gowen_allostery,altman,lee}. For example, Pashine \textit{et al.}~\cite{nidhi_aging} show that by leaving quasi-two-dimensional disordered foam arrays to age under uniform compression, they can be trained to exhibit a negative Poisson’s ratio, in which compression along one axis produces contraction (rather than expansion) along the transverse direction. This training method referred to as directed aging, benefits from a material’s ability to adapt over time to externally induced stress by lowering the energy cost of the imposed deformation~\cite{nidhi_aging}. Negative Poisson’s ratio materials are rarely found in nature. Thus, the ability to tune such a material response without intensive engineering demonstrates the power of training as a design technique. 

Our metamaterials are disordered arrays fabricated out of thin sheets of liquid crystal elastomer. Liquid crystal elastomers are elastomeric polymer networks that contain covalently bound liquid crystals (mesogens)~\cite{degennes,finkelmann,warner}. In a polydomain LCE, the liquid crystalline phases form locally ordered microscopic domains within the bulk material. The manipulation of the domain alignment leads to a host of exceptional material properties including stimuli-responsive mechanical and optical transformations, enhanced stress dissipation, and “soft elasticity” so that the material endures large strains with little increase in the stress as the mesogens rotate in the direction of the applied strain~\cite{warner,jin,sticky,luo,photoswitchable}. The mesogen alignment can be controlled by the imposition of an external electrical, magnetic, or mechanical field~\cite{herbert,jiang}. Altering that alignment with mechanical strain produces a large number of configurations in these materials that can be accessed by simple inputs. This is advantageous in developing materials that can be trained.  Additionally, liquid crystal elastomers exhibit shape memory behavior~\cite{herbert} and by simply heating the material above its nematic to isotropic transition temperature, the original network configuration can be recovered. This enables the erasure of the trained functionality as desired. 

We demonstrate that disordered arrays of liquid crystal elastomers can be trained via directed aging  to exhibit a negative Poisson’s ratio; we then examine the material components that enable this adaptive behavior. Finally, we demonstrate the ability to reset these materials as a means of forgetting to retrain for a different function, in this case an allosteric-inspired mechanical behavior~\cite{hexner_creep,gowen_allostery}. These findings demonstrate a new class of trainable pluripotent metamaterials.

\section{Results}

\subsection*{Training for Negative Poisson's Ratio}
\begin{figure}[tbhp]
\centering
\includegraphics[width=3.375in, height=7.1in]{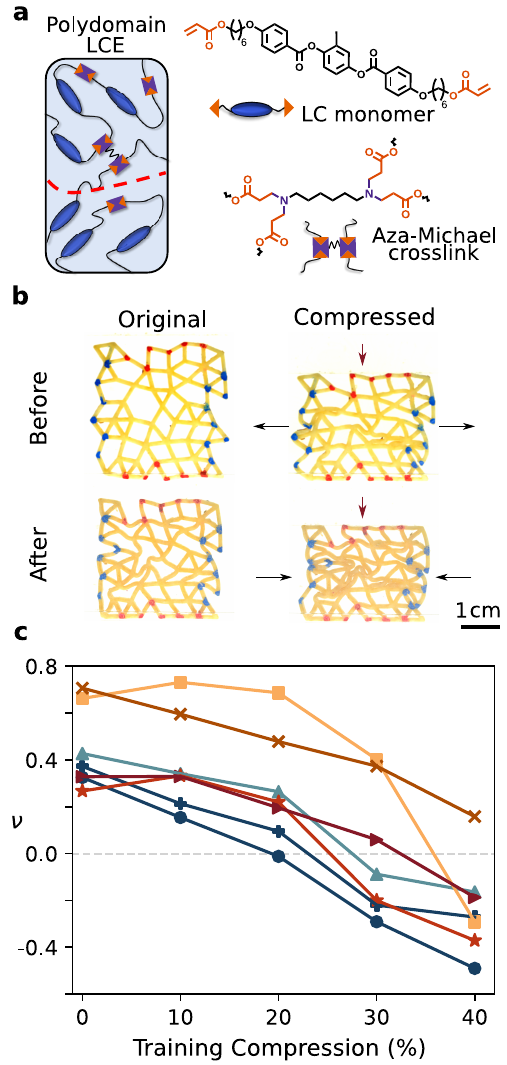}
\caption{Training disordered arrays of liquid crystal elastomers (LCEs) for auxetic behavior. \textbf{(a)} A schematic of microstructure in a polydomain LCE. \textbf{(b)} A sample LCE array is shown before (top row) and after aging under 40\% compression (bottom row). Before aging, the sample has $\nu > 0$ so that it expands horizontally under uniaxial vertical compression. After aging the sample contracts along the horizontal when uniaxial strain is applied, demonstrating an auxetic response, $\nu < 0 $. Blue and red dots are used for tracking strain at the boundary.  \textbf{(c)} The Poisson’s ratio, $\nu$, versus aging compression for six sample array geometries indicated by different symbols and colors. (Images of the arrays are shown in the SI.) $\nu$ drops significantly for aging compressions of 30-40\%. The array imaged in (b) is shown by duplicate experiments in dark blue (o and + signs).  }
\label{fig:fig_1}
\end{figure}

The polydomain liquid crystal elastomers (Fig.~\ref{fig:fig_1}a) in this study are synthesized using Aza-Michael chemistry, aiming for a molecular weight of 2000 g/mol between crosslinks~\cite{elina}. After synthesis, the LCE gels are washed, dried, and pressed into sheets approximately 2 mm thick with a lateral dimension of 45 mm.  Using a laser cutter, disordered array geometries are cut from these sheets (Fig.~S1). To examine the trainability of these metamaterials, we first train them to induce a negative Poisson’s ratio (auxetic) response. 

As initially fabricated, the arrays have a positive Poisson’s ratio so that when they are compressed uniaxially, they  expand in the transverse directions. This is shown by the first row of images in Fig.~\ref{fig:fig_1}b. This response is typical of nearly all natural and most synthetic materials. We use directed aging as a training protocol to modify the Poisson’s ratio: arrays are confined to an acrylic box and left to adapt under compression for 24 hours in an oven at 80°C. Temperature is incorporated to accelerate material aging, but aging can still be achieved at lower temperatures (Fig.~S2).  The arrays are left to cool under compression for at least 30 minutes in order to return to room temperature.  After cooling and removing the metamaterials from the boxes, they are stretched to their original size and the elastic response is remeasured. This process is repeated on the same sample for training compressions of increasing magnitude from approximately 10\% to 40\% by area. The material preparation and experimental setup are discussed in detail in \textit{Materials and Methods}.

Figure~\ref{fig:fig_1}c demonstrates that the Poisson’s ratio for liquid-crystal-elastomer disordered arrays can be tuned to have auxetic behavior when subjected to sufficiently large training compression. The Poisson’s ratio, $\nu$, is measured under a 3\% uniaxial compression as a function of training compression for six different disordered geometries (Fig.~S1). Initially, each sample has a positive Poisson’s ratio, $\nu > 0$; $\nu$ decreases as training is applied  at increasingly large compressions. Most arrays attain $\nu < 0$ when subjected to compressions above 30\%.

The magnitude of the change in $\nu$ varies between arrays with different disordered geometries. Those with larger void fractions tend to reach lower values of $\nu$. On the other hand, arrays with large void fractions but too few struts and nodes are dominated by boundary effects. The amount of void space in the array also dictates the maximum limit to the training compression. It becomes difficult to confine arrays at compressions greater than 40\% and forcing such confinement can lead to strut breakage as discussed below.

\subsection*{Adaptability of material components}
The ability of liquid crystal elastomer arrays to be trained implies that the material adapts under strain. The training might be accomplished through reconfiguration or weakening of the polymer network, or by rearrangement of the liquid crystals. To investigate the contributions of specific chemical components for producing adaptive behavior, we also studied the effects of training on non-liquid-crystalline amorphous polymer networks. 

\begin{figure}
\centering
\includegraphics[width=3.375in, height=2.5in]{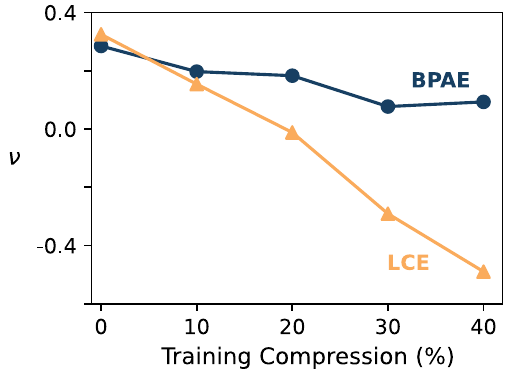}
\caption{ Training with variable sample chemistry. Here the training for negative Poisson’s ratio as a function of aging compression is compared between two samples with the same disordered array geometry (indicated by the circle in Fig.~S1), but a slightly different chemical composition. The sample with liquid crystal components (LCE) becomes auxetic at large training compressions while less change is seen in the non-liquid crystalline elastomer sample (BPAE).  }
\label{fig:f_2}
\end{figure}

To do this, we synthesized a bisphenol-A based elastomer (BPAE) with a similar crosslinking density to the LCEs. Each sample is trained under uniform compression for negative Poisson’s ratio behavior as described above.  Figure~\ref{fig:f_2} compares the Poisson’s ratio of an LCE array with an amorphous BPAE array at different training strains. The non-liquid crystalline BPAE array shows only a slight decrease in Poisson’s ratio as a function of training compression. Similar measurements on BPAE arrays with different geometries show little to no change in Poisson’s ratio for these non-liquid crystalline systems. Additionally, BPAE samples trained at compressions $\geq$ 20\% show a significant occurrence of strut fracture. 

For samples containing liquid crystals, a significant decrease in Poisson’s ratio is observed as a function of the training compression. We conclude that the alignment of the liquid crystals is the essential material adaptation that allows for a trainable elastic response in these metamaterials.

\subsection*{“Forgetting” through thermal reset}

An important additional consequence of liquid crystal alignment is further demonstrated by the ability of these LCE arrays to be “reset”. Originally, the locally ordered liquid crystalline domains are isotropically oriented throughout the sample.  Polydomain LCE-based materials have been shown to exhibit enhanced stress dissipation under compression, enabled by the realignment of the liquid crystalline domains~\cite{agrawal,song,jeon}.

When LCE metamaterials are compressed, stress is distributed non-uniformly throughout the sample; some struts are compressed while others are stretched, leading to a distribution of aligned domains. Once the arrays have been trained via compression and then strained back to their original size, they relax back to a size and geometry that is different from the original (Fig.~\ref{fig:fig_1}b).  This change in geometry is important for modifying the arrays’ properties, and is sustained over timescales much longer than the training time (Fig.~S3) However, upon heating the samples above the liquid crystal transition temperature (Fig.~S4), the arrays return almost completely to their initial configuration due to the shape memory effects of the LCE.  

The ability of these liquid crystal elastomer arrays to hold their trained state over long periods and then rapidly return to their initial geometry  once they are reset is a feature that has not been observed in previously reported polymer arrays~\cite{nidhi_aging,gowen_allostery}.  This introduces the exciting prospect of a material that can be re-trained. 

\begin{figure}
\centering
\includegraphics[width=3.375in, height=3in]{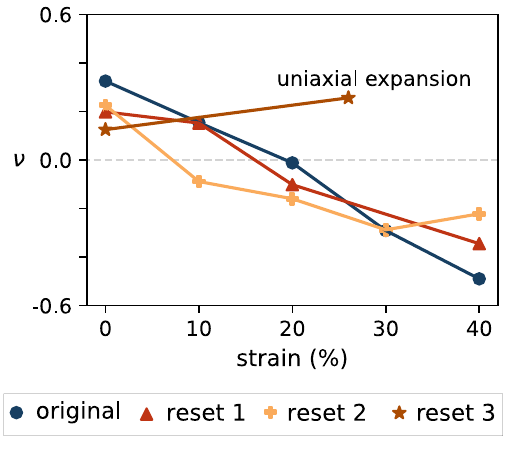}
\caption{  Thermal reset of array geometry. One sample is trained for negative Poisson’s ratio (circles) and then reset by heating above the liquid crystal transition temperature (Fig.~S1). The reset geometry is retrained using the same methods and closely follows the initial trend (triangles). The array is again reset by a combination of heating and light stretching and retrained a third time (plus signs). The three data sets show very similar behavior.  The array is reset a fourth time and aged under uniaxial expansion (stars). This results in an increased Poisson’s ratio.  Despite sequential aging and resetting the material is robustly able to “forget” prior training to be tuned again.   }
\label{fig:fig_3}
\end{figure}

We demonstrate this by taking an array that was trained incrementally under compression from 10-40\%, resetting it through applied heat, and then retraining it for the same auxetic functionality as before. While it becomes increasingly difficult to reset the arrays by heat alone after the first training cycle, the application of light tension to the array while heating allows it to return very close to its initial geometric configuration.  The results from the retrained arrays are shown by the triangle and plus symbols in Fig.~\ref{fig:fig_3}. Aside from a slight decrease in the initial Poisson’s ratio, the array’s response replicates almost exactly the results from the first training run (circle symbols) of $\nu$ versus training compression.

For the final reset, instead of again training in the same way under compression, we train the array by applying a uniaxial tension of about 26\% to the sample, as shown by the star symbols in Fig.~\ref{fig:fig_3}. This procedure leads to an increase in the Poisson’s ratio, demonstrating that the reset material is robust to retraining for other functions as well (in this case, tuning for an increased Poisson‘s ratio). 

As emphasized above, the ability of these material arrays to “forget” previous training also introduces the prospect of creating pluripotent materials~\cite{martin}. We demonstrate this by resetting one of the auxetic-trained arrays and then training it for a new type of function, inspired by allosteric interactions in proteins: this training modifies local interactions between nodes and struts in the array. 

\begin{figure}[t]
\centering
\includegraphics[width=3.375in, height=4.75in]{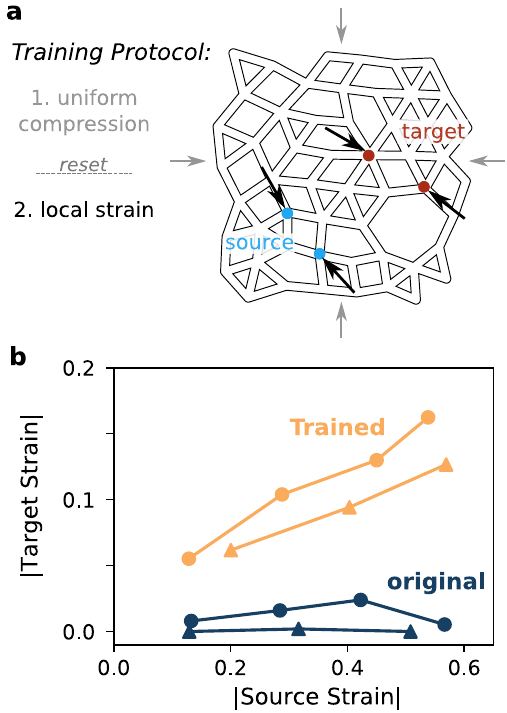}
\caption{  Training for local mechanical function. \textbf{(a)} The diagram demonstrates how arrays are initially trained for negative Poisson’s ratio under compression (gray arrows) and then reset and trained to couple locally a set of source (light blue) and target (red) nodes by straining them (black arrows). \textbf{(b)} The absolute value of the output target strains as a function of the absolute value of the input source strain for two retrained array geometries (circles and triangles) is measured. When the source struts are compressed, the target struts initially have little to no response. After 48 hours of training, the target responds with a strain of almost 20\% at large source strains indicating that the source and target sites have become coupled.    }
\label{fig:fig_4}
\end{figure}

For the allosteric-inspired function, a pair of source nodes and target nodes are chosen such that when strain is applied to the two source nodes, hardly any displacement is observed at the two target nodes. Training is implemented by simultaneously applying strain to both the source and target nodes, as shown in Fig.~\ref{fig:fig_4}a, and leaving them in this configuration at 80°C for 48 hours. The arrays are then cooled back to ambient temperature, and the nodes are stretched slightly before measurement. Figure~\ref{fig:fig_4}b shows the measured target strain when strain is applied to the source nodes for two retrained arrays. Before training, the target strain remains approximately zero for a range of source strains. After  training there is a significant response at the target indicating that the bonds were successfully coupled. 

\section*{Discussion and Conclusions}

This work expands previously described training techniques to a new class of materials, which demonstrate not only training for one function but also the ability to be re-trained and to  develop new desired function, \textit{i.e.}, pluripotency. Liquid crystal elastomer arrays, despite being significantly stiffer than the EVA (poly(ethylene-co-vinyl acetate)) foam arrays described by Pashine \textit{et al.}~\cite{nidhi_aging}, can achieve negative Poisson’s ratio values using the same training procedures. 

These experiments also clarify some limits of this training technique. For an isotropic two-dimensional material, the Poisson’s ratio is related to the bulk ($B$) and shear ($G$) elastic moduli: $\nu = (B-G)/(B+G)$. Thus, as the bulk modulus becomes smaller than the shear modulus the material becomes increasingly auxetic. In the experiments described above, the training process of applying uniform compression to the arrays lowers $B$ leading to a decrease in the Poisson’s ratio.   Pashine \textit{et al.}~\cite{nidhi_aging} showed that auxetic foam arrays adapt through changes in geometry, including changes in strut-lengths and strut bending or buckling, and to a lesser extent to changes in the strut stiffness. 

 Training many polymer materials may be difficult or impossible to implement because the stresses imposed by bending or buckling often lead to brittle fracture of the strut. In a few cases, under large compressions or tensions, fracture of isolated struts was also observed in liquid crystal elastomer samples.   The success of both liquid crystal elastomers and EVA foams in training suggests that a general characteristic of trainable materials is their ability to undergo large strains without significant fracture through plastic deformation.

Unlike the EVA foams which undergo permanent plastic deformations during the aging process, this work demonstrates that liquid crystal elastomers can be functionally reset. Resetting the geometry allows the same array to be trained first for a bulk response, and later for a completely different local mechanical response, such as mechanical allostery. Thus, these LCE arrays exhibit a form of pluripotency. 

In conclusion, these experiments demonstrate that directed aging can be used to train in various types of mechanical functions in the LCE metamaterials.  Additionally, training with the LCE arrays introduces the benefit of “forgetful” materials leading to pluripotency through the reset mechanism and re-trainability demonstrated in this work.

\section*{Materials and Methods}

\subsection*{Materials}
2-Methyl-1,4-phenylene bis(4-((6-(acryloyloxy)hexyl)oxy)benzoate) (RM 82) (LC monomer) was purchased either from Daken Chemical Limited or from Wilshire Technologies. 4-Nitrophenol was purchased either from Fluka or from Sigma-Aldrich. All other reagents were purchased from Sigma-Aldrich and were used as received unless otherwise noted. All solvents were purchased from Fisher Scientific and were used as received unless otherwise noted.

\subsection*{Synthesis of LCE and non-LCE networks}
LCEs were synthesized using aza-Michael chemistry as previously described~\cite{elina} , where RM 82 (LC monomer) (7.40 g, 0.011 mol), hexamethylenediamine (0.470 g, 0.004 mol), hexylamine (0.293 g, 0.002 mol), and 4-nitrophenol (0.153 g, 0.0011 mol) were added in a vial with 10 ml of toluene. The reaction mixture was sonicated to improve dissolution, then purged with nitrogen gas for 15 minutes before curing it at 80 °C. After 16 hours, an LCE gel was obtained. BPAE gels were synthesized exactly as LCEs, except the LC monomer (RM 82) was replaced by the BPA based monomer (Bis(4-acryloxypolyethoxyphenyl)propane). 

Both LCE and BPAE gels were washed separately overnight with acetone by placing them in a Soxhlet apparatus. The gels were then dried in a vacuum oven at 60 °C overnight. The dried gels were pressed at 160 °C for an hour under 4 tons pressure to obtain thick films ($\approx$ 45~mm x 2~mm x 45~mm). Differential scanning calorimetry was performed to determine the thermal transition temperatures of both LCE and BPAE films (Fig.~S4 and Fig.~S5). 

The films were laser cut using a Universal Laser Systems Ultra X6000 into disordered array patterns based on 2D simulated jammed particle packings where nodes correspond to particle centers and struts correspond with particle contacts. 

\subsection*{Measurement and Training Protocol}

For measurement, arrays are mounted above a light box between two pieces of acrylic. This prevents buckling out of the plane during compression. A thin plunger is connected to an Applied Motion Products integrated stepper motor (STM 23S-2AE) to apply uniaxial strain by compressing the top surface of the array while the bottom surface is held stationary. Arrays are coated with baby powder to reduce friction with the top and bottom acrylic plates. Images are captured by a Nikon D7000 DSLR camera. 

The array edges are painted with small red and blue dots to track the horizontal and vertical boundaries as strain is applied. An average axial strain is computed by fitting horizontal lines to the top and bottom red dots, while the transverse strain is computed by fitting vertical lines to the left and right edges. The Poisson’s ratio is computed by taking the negative ratio of the transverse to axial strain. 

Arrays are cut to designs that are approximately 45~mm x 45~mm and trained under compression in acrylic boxes that compress the arrays by 10-40\% of this area in 10\% increments. Arrays are screwed into compression boxes and placed in a Fisher Scientific Vacuum Oven Model 280 in 24~hr increments at 80\textdegree C. Once removed from the oven, arrays are left to cool to room temperature under compression for at least 30~min. Temperature is used to effectively speed up the aging process. We measured similar effects in arrays aged at 50\textdegree  C (Fig.~S2), but with faster relaxation times- indicating that more training time would be required to access the same results at lower temperatures. 

Training under compression conditions the arrays to be compressed between their original size and the size of the aging box. When arrays finish aging, they have often adapted exactly to the box size. Thus, to measure aging effects the arrays are stretched back to their original size using 3D printed platforms in which the negative space of the array is elevated. Once stretched, the arrays are re-measured. 

\section*{Acknowledgments}

The authors would like to thank Severine Atis, Chuqiao Chen, Efi Efrati, Varda Hagh, and Heinrich Jaeger for helpful discussions and insights. They would also like to thank Nidhi Pashine for her contributions in advising the experimental design. This research was supported by the University of Chicago Materials Research Science and Engineering Center, NSF-MRSEC program under award NSF-DMR 2011854. Parts of this work were carried out at the Soft Matter Characterization Facility and at the Materials Research Science and Engineering Center at the University of Chicago (MRSEC NSF DMR-2011854).



\bibliography{main}

\end{document}